# Optical toroidal dipolar response by an asymmetric double-bar metamaterial

Zheng-Gao Dong,[1,2,a)] J. Zhu,[2] Junsuk Rho,[2] Jia-Qi Li,[1] Changgui Lu,[2] Xiaobo Yin,[2] and X. Zhang[2]

[1] Physics Department, Southeast University, Nanjing 211189, China

[2] 5130 Etcheverry Hall, Nanoscale Science and Engineering Center, University of California, Berkeley, California 94720-1740, USA

We demonstrate that the toroidal dipolar response can be realized in the optical regime by designing a feasible nanostructured metamaterial, comprising asymmetric double-bar magnetic resonators assembled into a toroid-like configuration. It is confirmed numerically that an optical toroidal dipolar moment dominates over other moments. This response is characterized by a strong confinement of an $E$-field component at the toroid center, oriented perpendicular to the $H$-vortex plane. The resonance-enhanced optical toroidal response can provide an experimental avenue for various interesting optical phenomena associated with the elusive toroidal moment.

A family of toroidal moments characterized by vortex distributions of magnetic dipoles is regarded as the third kind of electromagnetic moments, though traditionally not accounted in the common multipole expansion of polarization and magnetization.[1-4] Mathematically, it can be considered as toroidization (*T*) with its dipolar moment defined as $\vec{T} = \sum_i \vec{r}_i \times \vec{M}_i$, where $\vec{M}_i$ is the magnetic dipolar moment, $\vec{r}_i$ is the dipole displacement relative to the toroidal center, and *i* represents the *i*th magnetic dipole that contributes to an *H*-vortex ordering.[5-7] Because the toroidal dipolar moment violates the symmetries of both space-inversion and time-reversal operation, a lot of intriguing properties can be expected, such as the $\vec{T}$ induced nonreciprocal refraction,[8] dichroism,[9] second-harmonic generation with an enhanced nonlinear susceptibility,[10] and other magnetoelectric effects. In particle physics, the elusive $\vec{T}$, produced by the electron spin or orbital movement, is attracting more and more attentions.[2] However, this kind of moment was traditionally ignored in macroscopic materials, due to the fact that the permanent toroidal ordering in natural magnetic materials was rarely observed, especially under room temperatures.[11] Moreover, theoretically predicted ferro-toroidic effects in naturally occurring media were generally very weak as compared with the spontaneous ferroelectric and ferromagnetic effects.[12]

Artificially structured materials are usually promising to find a way out of the dilemma confronted by natural media. During the last decade, metamaterials have attracted significant interest in exploring various electromagnetic phenomena, existing or not in naturally occurring media, such as negative refraction,[13] super imaging,[14] cloaking,[15] electromagnetic black hole,[16] and electromagnetically induced transparency.[17-20] Inspired by aforementioned studies on the static (permanent) toroidal moments in natural magnetic media, dynamic toroidal dipole that produces electromagnetic fields is drawing attention to the metamaterial community.[9,21-24] Recently, T. Kaelberer *et al.* demonstrated a toroidal dipolar resonance in the microwave regime by an elaborate multifold split-ring metamolecule.[21] It was stressed that the toroidal dipolar moment could be greatly enhanced due to the high-$Q$ resonant characteristic, which was also confirmed in a multi-fold structure by closed double rings.[22] Subsequently, the toroidal metamaterial was theoretically studied in far-infrared wavelengths by scaling down the split-ring structure.[24] However, these ring-based toroidal metamaterials are hard for fabrications, which makes it unfeasible for experimental explorations of the elusive optical phenomena associated with the toroidal dipolar moment.

In this work, we propose an optical toroidal metamaterial by asymmetric double bars (DB) numerically as well as experimentally. For the split-ring-based toroidal metamaterial, the electromagnetic wave was incident from the lateral direction of the toroidal metamolecule. In contrast, this work will show that the toroidal dipolar response can be excited by a normally incident light just by slightly lowering the structural symmetry, a similar approach with the excitation way of a dark mode.[25] To be frank, considering that the toroidal dipolar moment is inherently with simultaneous symmetry violations of both space inversion and time reversal, it is straightforward for us to introduce a geometric asymmetry for the aim of space-inversion symmetry breaking, in addition to the time-reversal symmetry breaking as an intrinsic result of the magnetization.

As is well known, a pair of metallic bars can be regarded as an artificial "magnetic atom" with a nonvanishing magnetic dipolar moment ($\vec{M}$) even at optical wavelengths.[26]

Based on such an optical magnetic meta-atom, the toroidal metamaterial proposed in this paper is composed of six asymmetric DB resonators in one unit cell (here, "asymmetric" means different lengths of DB), as shown in Fig. 1(a). In order to obtain the toroidal response (characterized by a vortex distribution of magnetic dipoles), the asymmetric top and bottom bars are flipped between the left and right halves, so that there is an inversion center (toroidal center $O$) for the metamolecule but the spatial rotation symmetry is broken with respect to the $z$-axis. This reduced spatial symmetry takes a same strategy as introduced in the split-ring-based toroidal metamaterials, which plays a vital role in the generation of the toroidal dipolar response.[21,24] For experimental measurements, the sample shown in Fig. 1(b) was fabricated by an electron-beam lithography (EBL) system, the transmission spectrum was measured by a Fourier transform infrared (FTIR) spectroscopy. For numerical simulations, the commercial solver CST Microwave Studio was used. In the simulation, the silver was described by using the Drude model $\varepsilon(\omega) = \varepsilon_\infty - \omega_p^2/(\omega^2 + i\omega\gamma)$, where the high-frequency permittivity $\varepsilon_\infty = 6.0$, the plasma frequency $\omega_p = 1.37 \times 10^{16} \text{s}^{-1}$, and the collision frequency $\gamma = 2.04 \times 10^{14} \text{s}^{-1}$.[27]

Figure 2 shows the transmission spectra of the proposed optical toroidal structure, with the $x$-polarized incident waves impinging normally to the film plane (i.e., incident along the $z$-direction). The experimental and numerical spectra are in good agreement except that some of the hybrid resonant modes with low quality factors are lost in the FTIR measurement data due to a heavy metallic loss. Corresponding to the resonance around 1.24 μm, the toroidal response characterized by an $H$-field vortex is clearly observed, as shown in the inset of Fig. 2 (also see a supplementary multimedia file online). Note that such a toroidal ordering is a direct result of the asymmetric DB metamolecule, under the normal-to-surface incidence it can neither be found in the symmetric DB case ($s = 0$ nm), nor be found in the asymmetric non-flipped DB metamolecule. For a comparison, figure 3 shows the simulation spectra as well as the corresponding *H*-field patterns of symmetric DB and asymmetric non-flipped DB metamolecules. Two types of resonant mode are obvious, one with the same *H*-field distribution for the left and right symmetric halves of the structure (no vortex), the other with antidirectional *H* field localized within the V-shaped elements. It is easy to understand that there is no toroidal dipolar response for these two configurations since the phase of the normally impinging plane wave is synchronous with respect to the left and right mirror halves of these two metamolecules. Moreover, it should be emphasized that the normal-to-surface incident configuration is distinguished from laterally incident cases proposed in literatures,[21,22,24] this can greatly contribute to optical experiments associated with this toroidal dipolar response.

As is known, the dipolar toroidization due to the closed magnetic-dipole ordering is an electric-type toroidal response, represented by a confined enhancement of the *E* field oriented parallel to the rotational axis of the toroidal DB structure (*z*-axis). From the simulation results shown in Fig. 4, the *E* field probed at the toroidal center point (*O*) is dominated by its *z*-component for the toroidal resonance around 1.24 μm, rather than the *x*- or *y*- component. The inset of Fig. 4 shows clearly that six DB heads (i.e., inner DB ends) have their $E_z$ fields reinforced one another at the toroidal central area. The locally enhanced $E_z$ field, as a result of the toroidal dipolar response, is promising for situations such as nonlinear optical phenomena, optical nano-cavity, and nano-scale optical sensing. This field enhancement capability becomes even more evident if it is noticed that such a resonant mode is simultaneously tightly-concentrated for *E* and *H* fields in the toroidal central area and DB gaps, respectively.

To investigate the toroidal dipolar response quantitatively, decomposed scattering fields, in terms of the multipole scattering theory, are numerically calculated according to the volume current density distribution.[3,21,22,24] Here, a single metamolecule with open boundaries were used for simulations, in contrast to periodic boundaries in calculating the transmission spectrum shown in Fig. 2. It is found from Fig. 5 that a predominant toroidal dipolar moment $T_z$, corresponding to the resonant mode with an *H*-vortex pattern, is responsible for the scattering field around the wavelength of 1.24 μm. As far as the $T_z$ is concerned, it should not contribute to the far-field scattering (i.e., no corresponding resonant characteristic in the spectrum) since it oscillates normal to the plane of the metamaterial array. However, as demonstrated in literatures,[17,20,25] introduction of a slight symmetry-breaking design in the structure can make an otherwise inaccessible mode accessible in the far field (usually resulted in a resonant spectrum). Therefore, it is reasonable here that the torodial response ($T_z$) corresponds to a resonant dip around 1.24 μm. In addition, the polarization component $P_z$ can produce an *H*-field vortex as well as $T_z$ can. Nevertheless, the scattering power contributed by $P_z$ is only on a level of $10^{-32}$ (not shown in Fig. 5), and thus can not mask the toroidal response since it is 2 ~ 3 orders weaker than that contributed by $T_z$.

The toroidal dipolar moment achieved by this asymmetric DB structure, with inherent broken symmetries of both space inversion and time reversal, can provide a feasible platform to experimentally explore unique properties of this naturally elusive electromagnetic moment. However, the strength of the metamaterial-based optical toroidal dipolar moment, in terms of its scattering intensity, is confronted with a heavy optical loss due to the metallic elements. On the other hand, to explain the origination mechanism for the toroidal dipolar response, we can resort to the well-developed magnetic plasmon hybridization model,[28] as have been used for interpreting the split-ring-based toroidal metamaterials.[21,24]

To the last but not least, the geometric dependence of the toroidal dipolar resonance is presented in Fig. 6. Since the toroidal response comes from the hybridization effect of magnetic dipoles, it must be dependent on the magnetic resonance of an individual DB. Figure 6(a) shows that the toroidal response will move to shorter wavelengths with the increase of the DB gap $g$. In fact, the magnetic resonance in a DB structure can be distinct from the electric dipole of a single bar only when the DB gap is small enough to confine a strong magnetic field in the gap. On the contrary, increasing the inner radius $r$ leads to a red shift of the toroidal resonant wavelength, which is a result of the reduced coupling between DB elements [Fig. 6(b)]. In this work, in order to obtain a compact coupling between magnetic dipoles, six DB elements are used instead of four. As a matter of fact, according to the definition of a toroidal dipolar moment, the simplest configuration that can produce a toroidal dipolar moment comes from antiparallel magnetic dipoles.[7]

In summary, toroidal dipolar response is demonstrated in the optical regime by designing a toroid-like metamaterial comprising six DB elements with reduced symmetry. The *E*-field confinement characteristic at the toroidal center, as a consequence of the *H*-vortex distribution, is an essential signature of the toroidal moment that distinguishes it from an individual magnetic dipole. Such an electromagnetic concentration capability should be useful for situations such as nonlinear optical phenomena, optical nano-cavity, and nano-scale optical sensing. Moreover, it is expected that various theoretically predicted effects in earlier literatures, associated with the elusive toroidal moment, would be observed experimentally based on this optical toroidal structure.

This work was supported by the US National Science Foundation (NSF) Nanoscale Science and Engineering Center CMMI-0751621. We also acknowledge the National Natural Science Foundation of China (No.10904012, No. 11174051, No. 11004026, and No. 61007018) and Youth Study Plans by SEU and CSC. Junsuk Rho acknowledges a fellowship from the Samsung Scholarship Foundation, Republic of Korea.

FIG. 1. (Color online) The toroidal metamaterial composed of asymmetric DBs. The scales are $a = 240$ nm, $b = 50$ nm, $c = 25$ nm, $s = 50$ nm, $g = 50$ nm, $r = 80$ nm, and the unit scales $L_x = L_y = 760$ nm. The substrate is glass ($n_{glass} = 1.55$) and the spacer layer is silicon oxide ($n_{SiO2} = 1.45$). (a) The schematic of the proposed toroidal metamaterial. (b) The image of the fabricated sample taken by a scanning electron microscope, where the inset shows the detail of top and bottom bars with different lengths.

FIG. 2. (Color online) Transmission spectra of the toroidal structure from both experiment and simulation. The inset shows the *H*-vortex distribution at the toroidal resonant mode (enhanced online).

FIG. 3. (Color online) Comparison results of symmetric DB and asymmetric non-flipped DB metamolecules in terms of simulation spectra and *H*-field patterns corresponding to the resonances.

FIG. 4. (Color online) *E*-field components probed at the center of the toroidal structure as a function of the incident wavelength. The inset is a map of the $E_z$ field on the *O-xz* plane to show the predominant $E_z$-field concentration at the toroidal dipolar response (plotted at 1.24μm).

FIG. 5. (Color online) Decomposed scattering powers corresponding to different multipole moments.

FIG. 6. Geometric dependence of the toroidal dipolar resonance. (a) Dependence on the DB gap $g$. (b) Dependence on the inner radius $r$.

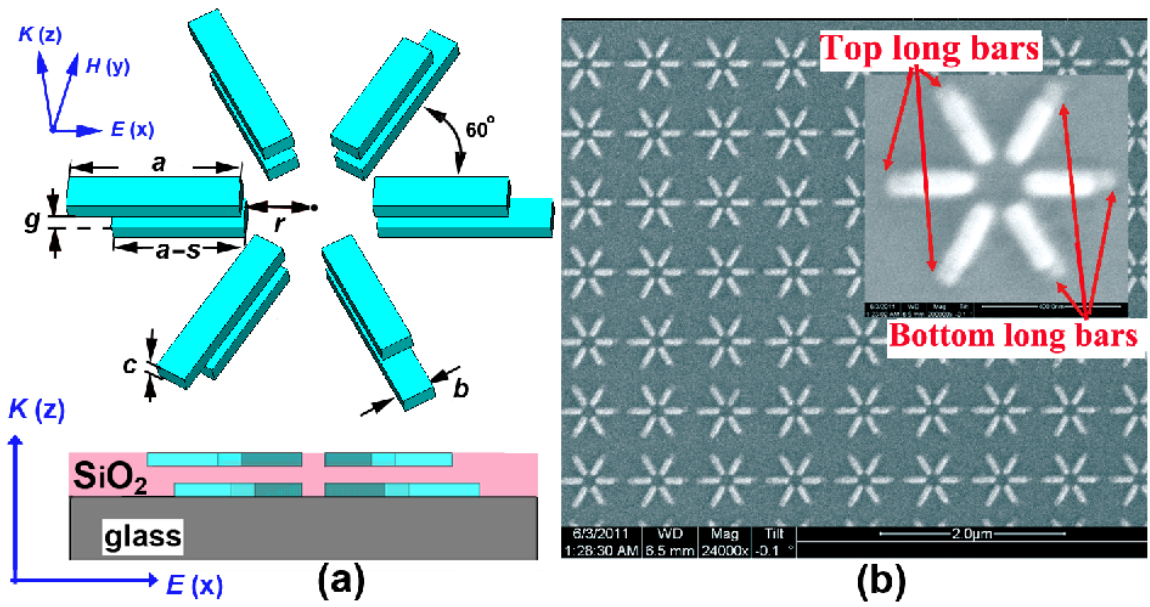

Fig. 1 Dong et al.

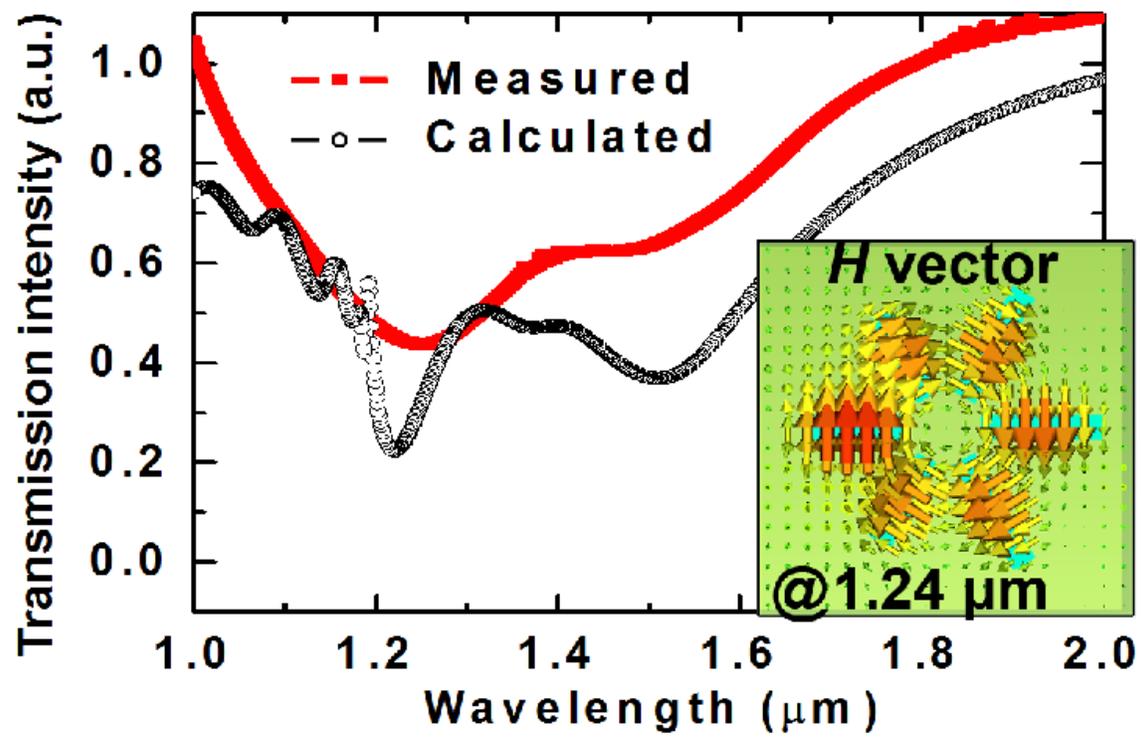

Fig. 2 Dong et al.

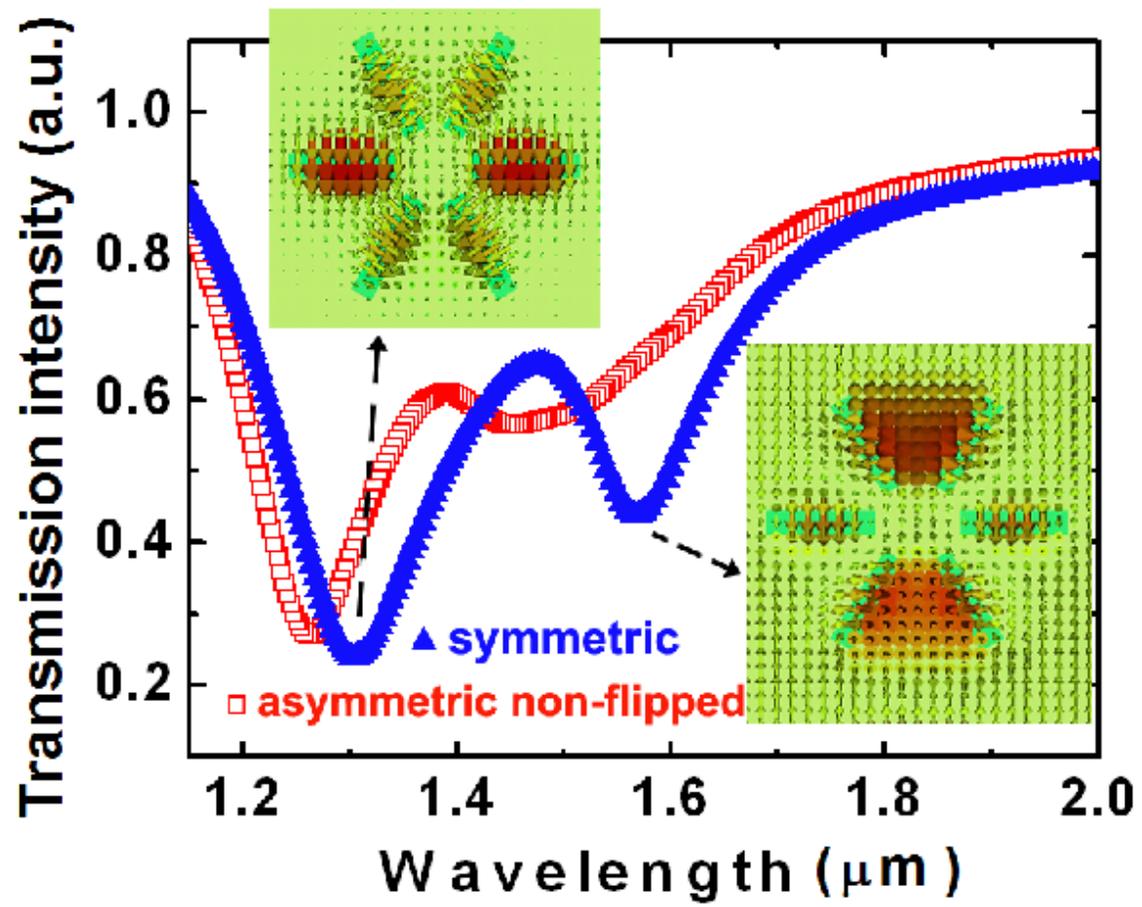

Fig. 3 Dong et al.

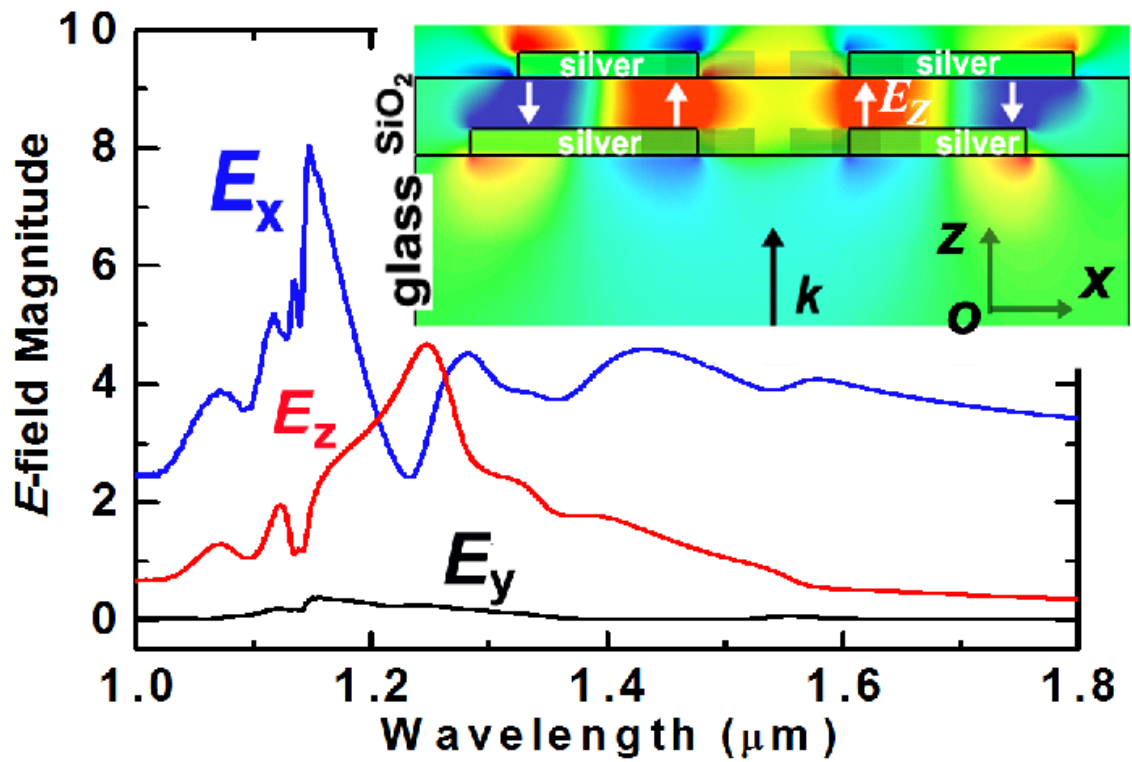

Fig. 4 Dong et al.

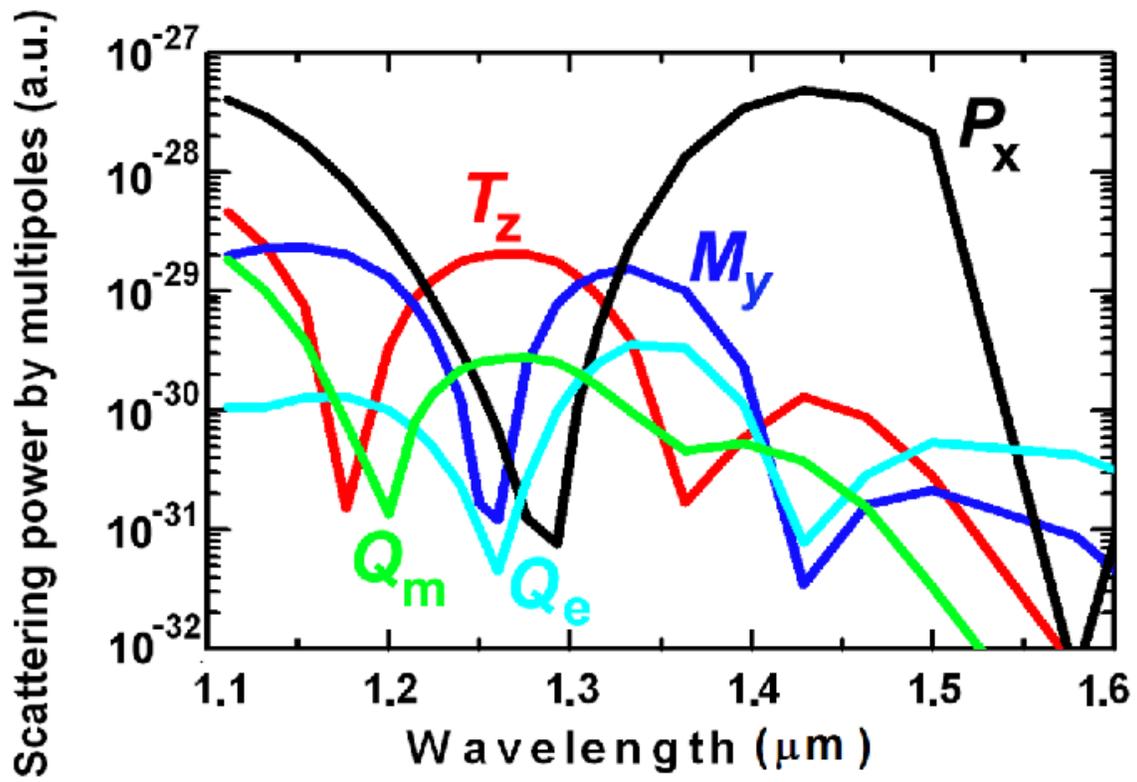

Fig. 5 Dong et al.

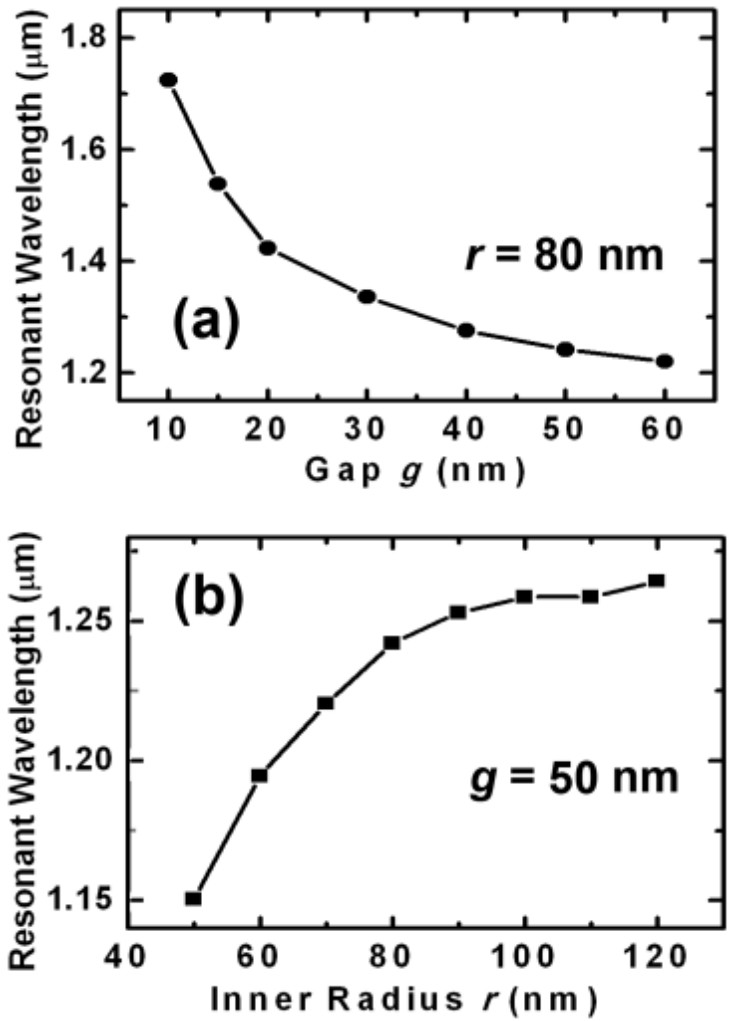

Fig. 6 Dong et al.

17